\documentclass[12pt, reqno]{amsart}

\usepackage[a4paper, margin=1.in, foot=.3in]{geometry}

\usepackage{graphicx, color}

\definecolor{darkblue}{rgb}{0,0,.5}
\definecolor{darkred}{rgb}{.5,0,0}
\definecolor{darkgreen}{rgb}{0,0.5,0}

\usepackage{hyperref}
\hypersetup {
    pdfstartview=FitH,
    colorlinks,
    citecolor=darkgreen,
    linkcolor=darkred,
    urlcolor=darkblue
}

\usepackage[noBBpl,slantedGreek]{mathpazo}

\usepackage{amsfonts,amssymb}
\allowdisplaybreaks
\numberwithin{equation}{section}

\makeatletter
\def\subsection{\@startsection{subsection}{2}%
  \z@{.5\linespacing\@plus.7\linespacing}{-.5em}%
  {\normalfont\bfseries\mathversion{bold}}}
\makeatother

\newcommand{\bbC}{\mathbb C}
\newcommand{\bbZ}{\mathbb Z}

\newcommand{\calC}{\mathcal C}
\newcommand{\calL}{\mathcal L}
\newcommand{\calM}{\mathcal M}
\newcommand{\calQ}{\mathcal Q}
\newcommand{\calR}{\mathcal R}
\newcommand{\calT}{\mathcal T}

\newcommand{\gothb}{\mathfrak b}
\newcommand{\gothg}{\mathfrak g}
\newcommand{\gothh}{\mathfrak h}
\newcommand{\gothsl}{\mathfrak{sl}}

\newcommand{\End}{\mathrm{End}}
\newcommand{\Hom}{\mathrm{Hom}}
\newcommand{\id}{\mathrm{id}}
\newcommand{\op}{\mathrm{op}}
\newcommand{\Osc}{\mathrm{Osc}}
\newcommand{\rme}{\mathrm e}
\newcommand{\rmi}{\mathrm i}
\newcommand{\tr}{\mathrm{tr}}

\newcommand{\sltwo}{\gothsl_2}
\newcommand{\uqsltwo}{U_q(\gothsl_2)}
\newcommand{\lsltwo}{\calL(\gothsl_2)}
\newcommand{\uqlsltwo}{U_q(\calL(\gothsl_2))}
\newcommand{\tlsltwo}{\widetilde{\calL}(\gothsl_2)}
\newcommand{\uqtlsltwo}{U_q(\widetilde{\calL}(\gothsl_2))}
\newcommand{\uqlbp}{U_q(\calL(\gothb_+))}

\title{Universal integrability objects}

\author[H. Boos]{Herman Boos}
\address{Fachbereich C -- Physik, Bergische Universit\"at Wuppertal, 42097 Wuppertal, Germany}
\email{boos@physik.uni-wuppertal.de}

\author[F. G\"ohmann]{Frank G\"ohmann}
\address{Fachbereich C -- Physik, Bergische Universit\"at Wuppertal, 42097 Wuppertal, Germany}
\email{goehmann@physik.uni-wuppertal.de}

\author[A. Kl\"umper]{Andreas Kl\"umper}
\address{Fachbereich C -- Physik, Bergische Universit\"at Wuppertal, 42097 Wuppertal, Germany}
\email{kluemper@uni-wuppertal.de}

\author[Kh. S. Nirov]{\vskip .2em Khazret S. Nirov}
\address{Institute for Nuclear Research of the Russian Academy of Sciences, 60th October Ave 7a,
117312 Moscow, Russia}
\curraddr{Fachbereich C -- Physik, Bergische Universit\"at Wuppertal, 42097 Wuppertal, Germany}
\email{knirov@physik.uni-wuppertal.de}

\author[A. V. Razumov]{Alexander V. Razumov}
\address{Institute for High Energy Physics, 142281 Protvino, Moscow region, Russia}
\curraddr{Max-Planck-Institut f\"ur Mathematik, Vivatsgasse, 7, 53111, Bonn, Germany}
\email{Alexander.Razumov@ihep.ru}

\begin{document}

\begin{abstract}
We discuss the main points of the quantum group approach in the theory of quantum integrable
systems and illustrate them for the case of the quantum group $\uqlsltwo$. We give a complete set
of the functional relations correcting inexactitudes of the previous considerations. A special
attention is given to the connection of the representations used to construct the universal
transfer operators and $Q$-operators.
\end{abstract}

\maketitle


\section{Introduction}

The modern approach to a wide class of quantum integrable systems is based on the concept of a
quantum group introduced by Drinfeld and Jimbo \cite{Dri87, Jim85}. Here all the objects describing
the model and related to its integrability are obtained from the universal $R$-matrix of the
underlying quantum group. For the first time this approach was used by Bazhanov, Lukyanov and
Zamolodchikov \cite{BazLukZam96, BazLukZam97, BazLukZam99}, see also the paper \cite{AntFei97}.

The universal $R$-matrix is an element of the tensor product of two copies of the quantum group
under consideration. The objects related to integrable systems are obtained by fixing
representations of the factors of the tensor product. By historical reasons, it is customary to
call the representation space of one of the factors the {\em auxiliary space\/}, and the
representation space of the other one the {\em quantum space\/}. For definiteness, we assume that
the auxiliary space is associated with the first factor, and the quantum space with the second one.
In fact, fixing the representation for the auxiliary space we define an object related to
integrability, while the choice of the representation for the quantum space defines a physical
model. For example, a square lattice vertex model and the related spin chain arise when we take for
the quantum space a tensor power of finite-dimensional representations of a quantum group. The
basic example here is the six-vertex model and the XXZ spin chain. If the quantum space is the
representation space of a certain infinite-dimensional vertex representation of the quantum group,
we have a two-dimensional quantum field theory.

If we fix the representation for the auxiliary space only, we obtain universal objects which do not
depend on the physical model. It appears that it is possible to derive for these objects the
universal functional relations responsible for the integrability, see, for example, \cite{AntFei97,
BazTsu08}. The functional relations for a concrete physical model can be obtained then by fixing
the representation of the quantum group in the quantum space.

In this talk we shortly discuss the main points of the quantum group approach and illustrate them
for the case of the quantum group $\uqlsltwo$. We give a complete set of the functional relations
correcting inexactitudes of the previous considerations. A special attention is given to the
connection of the representations used to construct the universal transfer operators and the
universal $Q$-operators. Additional details can be found in the paper \cite{BooGoeKluNirRaz12}.

\newpage

\section{Quantum group approach}

\subsection{General remarks}

\subsubsection{Quantum groups}

Let $\gothg$ be a Kac--Mody algebra \cite{Kac90}. A quantum group $U_q(\gothg)$ determined by
$\gothg$ is a Hopf algebra of a special type. As a Hopf algebra, it is supplied with an associative
multiplication with a unit, a coassociative comultiplication with a counit, and an antipode.

The quantum group $U_q(\gothg)$ can be considered as a 'deformation' of the enveloping algebra of
the Lie algebra $\gothg$. Depending on the sense of $q$, there are at least three definitions of a
quantum group. According to the first definition, $q = \exp \hbar$, where $\hbar$ is an
indeterminate, according to the second one, $q$ is an indeterminate, and according to the third
one, $q = \exp \hbar$, where $\hbar$ is a complex number such that $q \ne 0, \pm 1$. In the first
case a quantum group is a $\bbC[[\hbar]]$-algebra, in the second case a $\bbC(q)$-algebra, and in
the third case it is just a complex algebra. For our purposes, it seems that it is most convenient
to use the third definition. Therefore, we define a quantum group as a $\bbC$-algebra, see, for
example, the books \cite{JimMiw85, ChaPre94, EtiFreKir98}.

For any Hopf algebra $A$ with the comultiplication $\Delta$ we can define the opposite comultiplication:
\begin{equation*}
\Delta^{\mathrm{op}} = \Pi \circ \Delta,
\end{equation*}
where $\Pi$ is the element of $\End(A \otimes A)$ defined by the equation\footnote{In general, if
$A_1$ and $A_2$ are two algebras, we denote by $\Pi$ the element of $\Hom(A_1 \otimes A_2, A_2
\otimes A_1)$ defined by the equation $\Pi(a_1 \otimes a_2) = a_2 \otimes a_1$.}
\begin{equation*}
\Pi (a \otimes b) = b \otimes a.
\end{equation*}
A Hopf algebra $A$ is said to be {\em almost cocommutative\/}, if there exists an invertible
element $\calR \in A \otimes A$ such that
\begin{equation*}
\Delta^{\mathrm{op}}(a) = \calR \, \Delta(a) \, \calR^{-1}.
\end{equation*}
An almost cocommutative Hopf algebra $A$ is called {\em quasitriangular\/}, if
\begin{equation}
(\Delta \otimes \id) (\calR) = \calR^{13} \calR^{23}, \qquad
(\id \otimes \Delta) (\calR) = \calR^{13} \calR^{12}. \label{deltaid}
\end{equation}
In this case the element $\calR$ is called the {\em universal $R$-matrix\/}. The universal
$R$-matrix satisfies the Yang-Baxter equation for the universal $R$-matrix
\begin{equation}
\calR^{12} \, \calR^{13} \, \calR^{23} = \calR^{23} \, \calR^{13} \, \calR^{12}. \label{crcrcr}
\end{equation}
Any quantum group $U_q(\gothg)$ is a quasitriangular Hopf algebra.

\subsubsection{Spectral parameter}

Assume that a quasitriangular Hopf algebra $A$ is endowed with a family of automorphisms
$\Phi_\nu$, $\nu \in \bbC^\times$, satisfying the equation
\begin{equation}
\Phi_{\nu_1} \circ \Phi_{\nu_2} = \Phi_{\nu_1 \nu_2}. \label{phiphiphi}
\end{equation}
The spectral-parameter-dependent universal $R$-matrix is defined as
\begin{equation*}
\calR(\zeta_1 | \zeta_2) = (\Phi_{\zeta_1} \otimes \Phi_{\zeta_2}) \calR,
\end{equation*}
and the Yang--Baxter equation for the universal $R$-matrix (\ref{crcrcr}) gives the Yang--Baxter
equation for the spectral-parameter-dependent universal $R$-matrix,
\begin{equation}
\calR^{12}(\zeta_1 | \zeta_2) \, \calR^{13}(\zeta_1 | \zeta_3) \, \calR^{23}(\zeta_2 | \zeta_3) =
\calR^{23}(\zeta_2 | \zeta_3) \, \calR^{13}(\zeta_1 | \zeta_3) \, \calR^{12}(\zeta_1 | \zeta_2).
\label{crcrcrzeta}
\end{equation}

In the case when
\begin{equation}
(\Phi_\nu \otimes \Phi_\nu) \calR = \calR \label{phiphir}
\end{equation}
one has
\begin{equation*}
\calR(\zeta_1 \nu | \zeta_2 \nu) = \calR(\zeta_1 | \zeta_2)
\end{equation*}
for any $\nu \in \bbC^\times$. Here it is possible to define the universal $R$-matrix depending on
only one spectral parameter
\begin{equation*}
\calR(\zeta) = \calR(\zeta | 1).
\end{equation*}
Then one has
\begin{equation*}
\calR(\zeta_1 | \zeta_2) = \calR(\zeta_1^{\mathstrut} \zeta_2^{-1}),
\end{equation*}
and the spectral-parameter-dependent Yang--Baxter equation reads
\begin{equation*}
\calR^{12}(\zeta_1^{\mathstrut} \zeta_2^{-1}) \, \calR^{13}(\zeta_1^{\mathstrut} \zeta_3^{-1}) \,
\calR^{23}(\zeta_2^{\mathstrut} \zeta_3^{-1}) = \calR^{23}(\zeta_2^{\mathstrut} \zeta_3^{-1}) \,
\calR^{13}(\zeta_1^{\mathstrut} \zeta_3^{-1}) \, \calR^{12}(\zeta_1^{\mathstrut} \zeta_2^{-1}).
\end{equation*}

A simplest way to construct a family of automorphisms $\Phi_\nu$, satisfying equation
(\ref{phiphiphi}), is to assume that the Hopf algebra $A$ is endowed with a $\bbZ$-gradation,
\begin{equation*}
A = \bigoplus_{m \in \bbZ} A_m.
\end{equation*}
It is easy to see that the grading automorphisms
\begin{equation}
\Phi_\nu (a) = \sum_{m \in \bbZ} \nu^m a_m, \label{phinu}
\end{equation}
where $a = \sum_m a_m$, $a_m \in A_m$, satisfy equation (\ref{phiphiphi}).

\subsection{\texorpdfstring{Universal integrability objects from the universal
$R$-matrix}{Universal integrability objects from the universal R-matrix}}

\subsubsection{Preliminaries}

Let $U_q(\gothg)$ be a quantum group. Note that the universal $R$-matrix for $U_q(\gothg)$ is in
fact an element of $U_q(\gothb_+) \otimes U_q(\gothb_-) \subset U_q(\gothg) \otimes U_q(\gothg)$,
where $\gothb_+$ and $\gothb_-$ are the standard Borel subalgebras of $\gothg$ \cite{Ros89,
KirRes90, LevSoi90, TolKho92, KhoTol92}. Therefore, to construct universal integrability objects we
need representations of $U_q(\gothb_+)$. One can obtain representations of $U_q(\gothb_+)$ from
representations of  $U_q(\gothg)$ by the restriction, however, we need also representations which
cannot be obtained by this procedure.

Below $\varphi$ is a representation of $U_q(\gothg)$ in a vector space $V$, and $\rho$ is a
representation of $U_q(\gothb_+)$ in a vector space $W$ which cannot be extended to a
representation of the full quantum group $U_q(\gothg)$.

We assume that a family of automorphisms $\Phi_\nu$, $\nu \in \bbC^\times$, of $U_q(\gothg)$,
satisfying equation (\ref{phiphiphi}), is fixed and define the families of representations
\begin{equation*}
\varphi_\zeta = \varphi \circ \Phi_\zeta, \qquad \rho_\zeta = \rho \circ \Phi_\zeta,
\end{equation*}
parametrized by the spectral parameter $\zeta$.

\subsubsection{\texorpdfstring{$R$-operators and $R$-matrices}{R-operators and R-matrices}}

For any $\zeta_1, \zeta_2 \in \bbC^\times$ we define
\begin{equation*}
R_\varphi(\zeta_1 | \zeta_2) = (\varphi \otimes \varphi) (\calR(\zeta_1 | \zeta_2)) =
(\varphi_{\zeta_1} \otimes \varphi_{\zeta_2}) (\calR).
\end{equation*}
It is clear that $R_\varphi(\zeta_1 | \zeta_2)$ is an element of $\End(V) \otimes \End(V) \cong
\End(V \otimes V)$. We call it an $R$-operator. If $V = \bbC^k$ one can identify $R_\varphi(\zeta_1
| \zeta_2)$ with the corresponding $k^2 \times k^2$ matrix called an $R$-matrix. The Yang--Baxter
equation for the universal $R$-matrix (\ref{crcrcr}) or for the spectral-parameter-dependent
universal $R$-matrix (\ref{crcrcrzeta}) give the usual Yang--Baxter equation
\begin{equation*}
R_\varphi^{12}(\zeta_1 | \zeta_2) \, R_\varphi^{13}(\zeta_1 | \zeta_3) \, R_\varphi^{23}(\zeta_2 |
\zeta_3) = R_\varphi^{23}(\zeta_2 | \zeta_3) \, R_\varphi^{13}(\zeta_1 | \zeta_3) \,
R_\varphi^{12}(\zeta_1 | \zeta_2).
\end{equation*}
If equation (\ref{phiphir}) is valid, one can define the $R$-operator with one spectral parameter
\begin{equation*}
R_\varphi(\zeta) = R_\varphi(\zeta | 1)
\end{equation*}
which satisfies the Yang--Baxter equation of the form
\begin{equation*}
R_\varphi^{12}(\zeta_1^{\mathstrut} \zeta_2^{-1}) \, R_\varphi^{13}(\zeta_1^{\mathstrut}
\zeta_3^{-1}) \, R_\varphi^{23}(\zeta_2^{\mathstrut} \zeta_3^{-1}) =
R_\varphi^{23}(\zeta_2^{\mathstrut} \zeta_3^{-1}) \, R_\varphi^{13}(\zeta_1^{\mathstrut}
\zeta_3^{-1}) \, R_\varphi^{12}(\zeta_1^{\mathstrut} \zeta_2^{-1}).
\end{equation*}

One can also define an $R$-operator using two different representations of $U_q(\gothg)$, say
$\varphi_1: U_q(\gothg) \to \End(V_1)$ and $\varphi_2: U_q(\gothg) \to \End(V_2)$. In this case we
use the notation
\begin{equation*}
R_{\varphi_1, \varphi_2}(\zeta_1 | \zeta_2) = (\varphi_{1 \zeta_1} \otimes \varphi_{2 \zeta_2})(\calR).
\end{equation*}

It is clear that the operator $R_{\varphi_1, \varphi_2}(\zeta_1 | \zeta_2)$ is an element of
$\End(V_1) \otimes \End(V_2) \cong \End(V_1 \otimes V_2)$. It is useful to introduce the linear
mapping
\begin{equation*}
\check R_{\varphi_1, \varphi_2}(\zeta_1 | \zeta_2) = P \circ R_{\varphi_1, \varphi_2}(\zeta_1 | \zeta_2),
\end{equation*}
where the mapping $P$ is the element of $\Hom(V_1 \otimes V_2, V_2 \otimes V_1)$ defined by the equation
\begin{equation*}
P(v_1 \otimes v_2) = v_2 \otimes v_1.
\end{equation*}
The mapping $\check R_{\varphi_1, \varphi_2}(\zeta_1 | \zeta_2)$ is an element of $\Hom(V_1 \otimes
V_2, V_2 \otimes V_1)$ which serves as the intertwiner for the representations $\varphi_{1 \zeta_1}
\otimes_\Delta \varphi_{2 \zeta_2}$ and $\varphi_{2 \zeta_2} \otimes_\Delta \varphi_{1 \zeta_1}$ of
$U_q(\gothg)$ in the vector spaces $V_1 \otimes V_2$ and $V_2 \otimes V_1$
respectively.\footnote{We use the notation $\otimes_\Delta$ to distinguish between the tensor
product of representations and the usual tensor product of mappings, so that $(\varphi
\otimes_\Delta \psi)(a) = (\varphi \otimes \psi)(\Delta(a))$.} To show this, first write
\begin{equation*}
\varphi_{2 \zeta_2} \otimes \varphi_{1 \zeta_1} = \Pi \circ (\varphi_{1 \zeta_1} \otimes \varphi_{2
\zeta_2}) \circ \Pi.
\end{equation*}
Note that the first $\Pi$ at the right hand side of the above equation is an element of
$\Hom(\End(V_1) \otimes \End(V_2), \End(V_2) \otimes \End(V_1))$, and the second one is an element
of $\End(U_q(\gothg) \otimes U_q(\gothg))$. Since $U_q(\gothg)$ is an almost cocommutative Hopf
algebra, we come to the equation
\begin{equation*}
\Pi((\varphi_{2 \zeta_2} \otimes \varphi_{1 \zeta_1})(\Delta(a))) = (\varphi_{1 \zeta_1} \otimes
\varphi_{2 \zeta_2}) (\calR \, \Delta(a) \calR^{-1}).
\end{equation*}
Taking into account that
\begin{equation*}
\Pi(M_1 \otimes M_2) = P^{-1} \circ (M_1 \otimes M_2) \circ P
\end{equation*}
for any $M_1 \in \End(V_1)$ and $M_2 \in \End(V_2)$, we obtain
\begin{equation*}
(\varphi_{2 \zeta_2} \otimes_\Delta \varphi_{1 \zeta_1})(a) = \check R_{\varphi_1,
\varphi_2}(\zeta_1 | \zeta_2) \circ ((\varphi_{1 \zeta_1} \otimes_\Delta \varphi_{2 \zeta_2})(a))
\circ (\check R_{\varphi_1, \varphi_2}(\zeta_1 | \zeta_2))^{-1}.
\end{equation*}
Thus, the representations $\varphi_{1 \zeta_1} \otimes_\Delta \varphi_{2 \zeta_2}$ and $\varphi_{2
\zeta_2} \otimes_\Delta \varphi_{1 \zeta_1}$ are equivalent and the mapping $\check R_{\varphi_1,
\varphi_2}(\zeta_1 | \zeta_2)$ is the corresponding intertwiner.

The explicit forms of $R$-matrices were obtained from the corresponding universal $R$-matrices for
some representations of the quantum groups $U_q(\calL(\gothsl_2))$ \cite{KhoTol92, LevSoiStu93,
ZhaGou94, BraGouZhaDel94, BraGouZha95, BooGoeKluNirRaz10}, $U_q(\calL(\gothsl_3))$ \cite{ZhaGou94,
BraGouZhaDel94, BraGouZha95, BooGoeKluNirRaz10} and $U_q(\calL(\gothsl_3, \mu))$ \cite{KhoTol92,
BooGoeKluNirRaz11}, where $\mu$ is the standard diagram automorphism of $\gothsl_3$ of order 2.

\subsubsection{Universal monodromy operators and universal transfer operators} \label{s:ut}

We define a {\em universal monodromy operator\/} $\calM_\varphi(\zeta)$ by the equation
\begin{equation*}
\calM_\varphi(\zeta) = (\varphi_\zeta \otimes \id) (\calR).
\end{equation*}
It is clear that $\calM_\varphi(\zeta)$ is an element of the algebra $\End(V) \otimes U_q(\gothg)$.

The transfer operators are obtained via taking the trace over the representation space $V$ of the
representation $\varphi$ used to define the monodromy operators. We denote by $\tr$ the usual trace
on the algebra of endomorphisms under consideration.

In general, if $\tr_A$ is a linear mapping from an algebra $A$ to $\bbC$, satisfying the cyclic property
\begin{equation*}
\tr_A(a_1 a_2) = \tr_A(a_2 a_1),
\end{equation*}
we say that $\tr_A$ is a trace on $A$. It is useful to have in mind that a linear combination of
traces on an algebra $A$ is a trace on $A$.

If $\varphi$ is a representation of an algebra $A$ in a vector space $V$, we denote
\begin{equation*}
\tr_\varphi = \tr \circ \varphi.
\end{equation*}
It is evident that $\tr_\varphi$ is a trace on the algebra $A$. Due to the cyclic property of a
trace, if two representations $\varphi_1$ and $\varphi_2$ of an algebra $A$ are equivalent, then
$\tr_{\varphi_1} = \tr_{\varphi_2}$.

Let $t$ be a group-like element of $A$. This means that
\begin{equation}
\Delta(t) = t \otimes t. \label{deltat}
\end{equation}
Starting with the universal monodromy operator $\calM_\varphi(\zeta)$, we define the corresponding
{\em universal transfer operator\/} as
\begin{equation*}
\calT_\varphi(\zeta) = (\tr \otimes \id)(\calM_\varphi(\zeta)(\varphi_\zeta(t) \otimes 1)) =
(\tr_{\varphi_\zeta} \otimes \id) (\calR (t \otimes 1)).
\end{equation*}
It is common to call $t$ a {\em twist element\/}.

An important property of transfer operators is their commutativity. Let $\varphi_1: U_q(\gothg) \to
\End(V_1)$ and $\varphi_2: U_q(\gothg) \to \End(V_2)$ be two representations of $U_q(\gothg)$.
Using the definition of the universal transfer operator written as
\begin{equation*}
\calT_\varphi(\zeta) = (\tr_{\varphi_\zeta} \otimes \id) (\calR^{12} t^1),
\end{equation*}
we obtain
\begin{equation*}
\calT_{\varphi_1}(\zeta_1) \calT_{\varphi_2}(\zeta_2) = (\tr_{\varphi_{1 \zeta_1}} \otimes
\tr_{\varphi_{2 \zeta_2}} \otimes \id)(\calR^{13} \calR^{23} t^1 t^2).
\end{equation*}
Now rewriting equation (\ref{deltat}) as
\begin{equation*}
\Delta(t) = t^1 t^2
\end{equation*}
and having in mind (\ref{deltaid}), we see that
\begin{equation*}
(\Delta \otimes \id)(\calR (t \otimes 1)) = \calR^{13} \calR^{23} t^1 t^2.
\end{equation*}
Thus, we have
\begin{equation*}
\calT_{\varphi_1}(\zeta_1) \calT_{\varphi_2}(\zeta_2) = (\tr_{\varphi_{1 \zeta_1} \otimes_\Delta
\varphi_{2 \zeta_2}} \otimes \id)(\calR (t \otimes 1)).
\end{equation*}
In a similar way we determine that
\begin{equation*}
\calT_{\varphi_2}(\zeta_2) \calT_{\varphi_1}(\zeta_1) = (\tr_{\varphi_{2 \zeta_2} \otimes_\Delta
\varphi_{1 \zeta_1}} \otimes \id)(\calR (t \otimes 1)).
\end{equation*}
Since the representations $\varphi_{1 \zeta_1} \otimes_\Delta \varphi_{2 \zeta_2}$ and $\varphi_{2
\zeta_2} \otimes_\Delta \varphi_{1 \zeta_1}$ are equivalent, we have
\begin{equation}
\calT_{\varphi_1}(\zeta_1) \calT_{\varphi_2}(\zeta_2) = \calT_{\varphi_2}(\zeta_2)
\calT_{\varphi_1}(\zeta_1). \label{tzetatzeta}
\end{equation}

Let us prove one more useful property of the universal transfer matrices. Let $a$ be a group-like
element of $U_q(\gothg)$. Since
\begin{equation*}
\Delta(a) = \Delta^\op(a),
\end{equation*}
we have
\begin{equation*}
\calR^{12} a^1 a^2 = a^1 a^2 \calR^{12}.
\end{equation*}
If $a$ commutes with the twist element $t$, then
\begin{equation*}
\calR^{12} t^1 a^1 a^2 = a^1 a^2 \calR^{12} t^1.
\end{equation*}
Assuming that $a$ is an invertible element, we can rewrite this equation as
\begin{equation*}
(a^1)^{-1} ((\calR^{12} t^1) \, a^2) \, a^1 = a^2 (\calR^{12} t^1).
\end{equation*}
Now applying to both sides mapping $(\tr \circ \varphi_\zeta) \otimes \id$, we see that
\begin{equation*}
\calT_\varphi(\zeta) a = a \, \calT_\varphi(\zeta)
\end{equation*}
for any invertible group-like element $a \in U_q(\gothg)$ commuting with the twist element~$t$.

\subsubsection{\texorpdfstring{Universal $L$-operators and universal $Q$-operators}{Universal
L-operators and universal Q-operators}} \label{s:uq}

$L$-operators play in the construction of $Q$-operators the same role as monodromy operators in the
construction of transfer operators, and the definition of $L$-operators is very similar to the
definition of the monodromy operators. The main distinction here is that to define $L$-operators we
use the representation $\rho$ of $U_q(\gothb_+)$ which cannot be extended to a representation of
$U_q(\gothg)$. In fact, to obtain some useful objects one should choose representations $\rho$
defining $L$-operators and $Q$-operators to be related to representations $\varphi$ used to define
the monodromy operators and the corresponding transfer operators. Presently, we do not have a full
understanding of how to do it. It seems that the representation $\rho$ should be obtained from the
representation $\varphi$ via some limiting procedure, see \cite{BazHibKho02, HerJim11} and the
example below.

We define a {\em universal $L$-operator\/} by the equation
\begin{equation*}
\calL_\rho(\zeta) = (\rho_\zeta \otimes \id)(\calR).
\end{equation*}
It is clear that $\calL_\rho(\zeta)$ is an element of $\End(W) \otimes U_q(\gothb_-)$.

The corresponding {\em universal $Q$-operator\/} is defined by the relation
\begin{equation*}
\calQ_\rho(\zeta) = (\tr \otimes \id)(\calL_\varphi(\zeta)(\rho_\zeta(t) \otimes 1)) =
(\tr_{\rho_\zeta} \otimes \id) (\calR (t \otimes 1)).
\end{equation*}
Since $\check R_{\rho, \varphi}(\zeta_1, \zeta_2)$ is the intertwiner of the representations
$\rho_{\zeta_1} \otimes_\Delta \varphi_{\zeta_2}$ and $\varphi_{\zeta_2} \otimes_\Delta
\rho_{\zeta_1}$ they are equivalent. Therefore, one has
\begin{equation}
\calQ_\rho(\zeta_1) \calT_\varphi(\zeta_2) = \calT_\varphi(\zeta_2) \calQ_\rho(\zeta_1), \label{qzetatzeta}
\end{equation}
where we assume that the same twist element is used to define both the universal $Q$-operator and
the universal transfer matrix. As well as for the case of universal transfer operators, one can
show that any group-like element commuting with the twist element commutes with $Q$-operators
defined with the help of this twist element.

Using only the definition of $\calQ_\rho(\zeta)$, one can not prove the commutativity of
$\calQ_\rho(\zeta)$ for different values of the spectral parameter because $\rho$ cannot be
extended to a representation of the whole algebra~$U_q(\gothg)$ and the corresponding intertwiner
cannot be constructed in a direct way. However, as for the case of universal transfer matrices,
also here we can obtain the equation
\begin{equation}
\calQ_{\rho_1}(\zeta_1) \calQ_{\rho_2}(\zeta_2) = (\tr_{\rho_{1 \zeta_1} \otimes_\Delta \rho_{2
\zeta_2}} \otimes \id)(\calR (t \otimes 1)) \label{qzetaqzeta}
\end{equation}
valid for any representations $\rho_1$ and $\rho_2$ of $U_q(\gothb_+)$. Analysing the tensor
product of the representations $\rho_1$ and $\rho_2$ one obtains information about the product of
the operators $\calQ_{\rho_1}(\zeta_1)$ and $\calQ_{\rho_2}(\zeta_2)$. In this way one can prove
the functional relations.

\section{Example. Universal integrability objects}

We consider the example of the quantum group $\uqlsltwo$. The necessary representations of
$\uqlsltwo$ and of the corresponding Borel subalgebra can be constructed by using the homomorphisms
to the quantum group $\uqsltwo$ and to the $q$-oscillator algebra $\Osc_q$ respectively. Therefore,
we start with a discussion of the simplest representations of these algebras.

\newpage

\subsection{\texorpdfstring{Quantum group $\uqsltwo$}{Quantum group Uq(sl2)}}

\subsubsection{Definition}

Let $\hbar$ be a complex number such that $q = \exp \hbar$ is not equal to $0$ and~$\pm 1$. We
assume that $q^\nu$, $\nu \in \bbC$, means the complex number $\exp (\hbar \nu)$. The quantum group
$\uqsltwo$ is a unital associative $\bbC$-algebra generated by the elements $E$, $F$, and $q^{\nu
H}$, $\nu \in \bbC$, with the following defining relations
\begin{gather}
q^0 = 1, \qquad q^{\nu_1 H} q^{\nu_2 H} = q^{(\nu_1 + \nu_2)H}, \label{hhsl} \\
q^{\nu H} E q^{-\nu H} = q^{2 \nu} E, \qquad q^{\nu H} F q^{- \nu H} = q^{- 2 \nu} F, \label{hehfsl} \\
[E, F] = \kappa_q^{-1} (q^H - q^{-H}). \label{efsl}
\end{gather}
Here and below $\kappa_q = q - q^{-1}$. Note that $q^{\nu H}$ is just a notation, there is no an
element $H \in \uqsltwo$. In fact, it is constructive to identify $H$ with the standard Cartan
element of the Lie algebra $\gothsl_2$, and $\nu H$ with a general element of its Cartan subalgebra
$\gothh = \bbC H$. Using this interpretation, one can say that $q^{\nu H}$ is a set of generators
parameterized by the elements of the standard Cartan subalgebra of $\gothsl_2$.

The quantum group $\uqsltwo$ is also a Hopf algebra with the comultiplication
\begin{gather*}
\Delta(q^{\nu H}) = q^{\nu H} \otimes q^{\nu H}, \\*
\Delta(E) = E \otimes 1 + q^{-H} \otimes E, \qquad \Delta(F) = F \otimes q^H + 1 \otimes F,
\end{gather*}
and the correspondingly defined counit and antipode.

The monomials $E^i F^j q^{\nu H}$ for $i, j \in \bbZ_{\ge 0}$ and $\nu \in \bbC$ form a basis of
$\uqsltwo$. There is one more basis defined with the help of the {\em quantum Casimir element\/}
$C$ which has the form
\begin{equation*}
C = E F + \kappa_q^{-2} (q^{H - 1} + q^{- H + 1}) = F E + \kappa_q^{-2} (q^{H + 1} + q^{- H - 1}).
\end{equation*}
Here and below we use the notation $q^{\nu H + \mu} = q^\mu q^{\nu H}$, $\nu, \mu \in \bbC$. One
can verify that $C$ belongs to the center of $\uqsltwo$. It is clear that the monomials of the form
$E^{i + 1} C^j q^{\nu H}$, $F^{i + 1} C^j q^{\nu H}$ and $C^j q^{\nu H}$ for $i, j \in \bbZ_{\ge
0}$ and $\nu \in \bbC$ also form a basis of $\uqsltwo$.

\subsubsection{Simplest modules and representations} \label{sltwom}

The simplest $\uqsltwo$-modules have a basis consisting of eigenvectors of the operators
corresponding to the elements $q^{\nu H}$. Let $v$ be such a vector. It follows from  (\ref{hhsl})
that\footnote{Here and below considering representations we use the module notation.}
\begin{equation*}
q^{\nu H} v = q^{\nu \mu} v
\end{equation*}
for some $\mu \in \bbC$. The number $\mu$ is called a {\em weight\/}, and the vector $v$ a {\em
weight vector\/} of weight~$\mu$.

The defining relations (\ref{hehfsl}) immediately give
\begin{equation*}
q^{\nu H} E v = q^{\nu (\mu + 2)} E v, \qquad q^{\nu H} F v = q^{\nu (\mu - 2)} F v.
\end{equation*}
These relations suggest us to consider a free vector space generated by the vectors $v_n$, $n \in
\bbZ$, such that
\begin{equation*}
F v_n = v_{n + 1}.
\end{equation*}
If $q^{\nu H} v_0 = q^{\nu \mu} v_0$, then we have
\begin{equation*}
q^{\nu H} v_n = q^{\nu (\mu - 2 n)} v_n.
\end{equation*}
As for the action of $E$ on $v_n$, it is natural to assume that
\begin{equation*}
E v_n = c_n v_{n - 1}
\end{equation*}
for some complex constants $c_n$. Now the defining relation (\ref{efsl}) gives
\begin{equation*}
c_{n + 1} = c_n + [\mu - 2 n]_q,
\end{equation*}
and, therefore,
\begin{equation*}
c_n = \lambda + [n]_q [\mu - n + 1]_q
\end{equation*}
for some constant $\lambda \in \bbC$. Here and below
\begin{equation*}
[\nu]_q = \kappa_q^{-1} (q^\nu - q^{-\nu}) = \frac{q^\nu - q^{-\nu}}{q - q^{-1}}
\end{equation*}
for any $\nu \in \bbC$.

We obtain a $\uqsltwo$-module determined by the equations
\begin{gather}
q^{\nu H} v_n = q^{\nu(\mu - 2 n)} v_n, \label{rIa} \\
E v_n = (\lambda + [n]_q [\mu - n + 1]_q) v_{n -1}, \qquad F v_n = v_{n + 1}. \label{rIb}
\end{gather}
We denote this module by $\widetilde V^{\mu, \lambda}$ and the corresponding representation by
$\widetilde \pi^{\mu, \lambda}$. The action of the quantum Casimir operator on the vectors of the
module $\widetilde V^{\mu, \lambda}$ is
\begin{equation*}
C v = \kappa_q^{-2} (\lambda + q^{\mu + 1} + q^{- \mu - 1}) v.
\end{equation*}

Introduce for the $\uqsltwo$-module $\widetilde V^{\mu, \lambda}$ a new basis formed by the vectors
\begin{equation*}
u_n = v_{n + k}
\end{equation*}
for some $k \in \bbZ$. Simple calculations give
\begin{gather*}
q^{\nu H} u_n = q^{\nu(\mu - 2 k - 2 n)} u_n, \\
E u_n = (\lambda + [k]_q [\mu - k + 1]_q + [n]_q [\mu - 2 k - n + 1]_q) u_{n -1}, \qquad F u_n = u_{n + 1}.
\end{gather*}
Thus the modules $\widetilde V^{\mu, \lambda}$ with
\begin{equation*}
\lambda = \lambda_0 + [k]_q [\mu_0 - k + 1]_q, \qquad \mu = \mu_0 - 2 k
\end{equation*}
are isomorphic for all $k \in \bbZ$ and fixed $\mu_0$ and $\lambda_0$.

\subsubsection{Highest weight modules}

Consider the $\uqsltwo$-module $\widetilde V^{\mu, \lambda}$ and assume that
\begin{equation*}
\lambda + [n]_q [\mu - n + 1]_q = 0
\end{equation*}
for some $n \in \bbZ$. Shifting the basis we see that up to an isomorphism we can assume that this
equation is valid for $n = 0$, so that $\lambda = 0$. Hence, we have
\begin{equation*}
E v_0 = 0.
\end{equation*}
It is clear that the vectors $v_n$, $n \in \bbZ_{\ge 0}$, form a basis of a $\uqsltwo$-submodule.
We denote it by $\widetilde V^\mu$ and the corresponding quotient representation by $\widetilde
\pi^\mu$ . The module $\widetilde V^\mu$ is a {\em highest weight module\/} with highest weight
$\mu$.

If $\mu$ equals a non-negative integer $m$, the linear hull of the vectors $v_n$ with $n > m$ is a
$\uqsltwo$-submodule of $\widetilde V^m$ isomorphic to the module $\widetilde V^{- m - 2}$. We
denote the corresponding finite-dimensional quotient module $\widetilde V^m / \widetilde V^{-m -
2}$ by $V^m$ and the corresponding quotient representation by $\pi^{m}$.

\subsubsection{Traces}

The trace defined by a representation $\widetilde \pi^{\mu, \lambda}$ for a general $\lambda$ is
singular. However, for $\lambda = 0$, using the representation $\widetilde \pi^\mu$ and denoting
\begin{equation*}
\widetilde \tr_\mu = \tr_{\widetilde \pi^\mu},
\end{equation*}
 we obtain that
\begin{equation*}
\widetilde \tr_\mu (E^{i + 1} C^j q^{\nu H}) = 0, \qquad \widetilde \tr_\mu(F^{i + 1} C^j q^{\nu H}) = 0,
\end{equation*}
and that
\begin{equation*}
\widetilde \tr_\mu(C^j q^{\nu H}) = \kappa_q^{-2 j} (q^{\mu + 1} + q^{- \mu - 1})^j \frac{q^{\nu \mu}}{1 - q^{-2 \nu}}
\end{equation*}
for $|q^{- 2 \nu}| < 1$. If $|q^{- 2 \nu}| > 1$ the trace of $C^j q^{\nu H}$ can be defined by analytic continuation.

Using the finite-dimensional representation $\pi^m$ and denoting
\begin{equation*}
\tr_m = \tr_{\pi^m},
\end{equation*}
we obtain
\begin{gather*}
\tr_m (E^{i + 1} C^j q^{\nu H}) = 0, \qquad \tr_m(F^{i + 1} C^j q^{\nu H}) = 0, \\*[.5em]
\tr_m(C^j q^{\nu H}) = \kappa_q^{-2 j} (q^{m + 1} + q^{- m - 1})^j \, [m + 1]_{q^\nu}.
\end{gather*}
One easily obtains the equation
\begin{equation}
\tr_m = \widetilde \tr_m - \widetilde \tr_{-m-2} \label{trm}
\end{equation}
which actually follows from the definition of the representation $\pi^m$.

One can define
\begin{equation}
\tr_\mu = \widetilde \tr_\mu - \widetilde \tr_{- \mu - 2} \label{trmu}
\end{equation}
for an arbitrary $\mu \in \bbC$. The mapping $\tr_\mu$ is a trace on $\uqsltwo$, however, it is not
generated by a representation of $\uqsltwo$.

\subsection{\texorpdfstring{$q$-oscillators}{q-oscillators}}

\subsubsection{Definition}

We start with reminding the necessary definitions, see, for example, the book \cite{KliSch97}. Let
$\hbar$ be a complex number such that $q = \exp \hbar \ne 0, \pm 1$. The $q$-oscillator algebra
$\Osc_q$ is a unital associative $\bbC$-algebra with generators $b^\dagger$, $b$, $q^{\nu N}$, $\nu
\in \bbC$, and relations
\begin{gather}
q^0 = 1, \qquad q^{\nu_1 N} q^{\nu_2 N} = q^{(\nu_1 + \nu_2)N}, \label{nn} \\
q^{\nu N} b^\dagger q^{-\nu N} = q^\nu b^\dagger, \qquad q^{\nu N} b \, q^{-\nu N} = q^{-\nu} b,
\label{nb} \\
b^\dagger b = \kappa_q^{-1} (q^N - q^{-N}), \qquad b \, b^\dagger
= \kappa_q^{-1} (q^{N + 1} - q^{- N - 1}). \label{bb}
\end{gather}
It is easy to understand that the monomials $(b^\dagger)^{i + 1} q^{\nu N}$, $b^{i + 1} q^{\nu N}$
and $q^{\nu N}$ for $i \in \bbZ_{\ge 0}$ and $\nu \in \bbC$ form a basis of $\Osc_q$.

\subsubsection{Simplest modules and representations} \label{oscm}

The simplest $\Osc_q$-modules have a basis consisting of eigenvectors of the operators
corresponding to the elements $q^{\nu N}$. If $v$ be such a vector, then it follows from (\ref{nn})
that
\begin{equation*}
q^{\nu N} v = q^{\nu \lambda} v
\end{equation*}
for some $\lambda \in \bbC$. In turn, the defining relations (\ref{nb}) give
\begin{equation*}
q^{\nu N} b^\dagger v = q^{\nu(\lambda + 1)} b^\dagger v,
\qquad q^{\nu N} b \, v = q^{\nu(\lambda - 1)} b \, v.
\end{equation*}
Having these equations in mind, let us consider a free vector space generated by the vectors $v_n$,
$n \in \bbZ$, and try to endow it with a structure of an $\Osc_q$-module assuming first that
\begin{equation*}
b^\dagger v_n = v_{n+1}.
\end{equation*}
Now, if we assume additionally that $q^{\nu N} v_0 = q^{\nu \lambda} v_0$, then
\begin{equation*}
q^{\nu N} v_n = q^{\nu (\lambda + n)} v_n,
\end{equation*}
and it is natural to expect that the action of $b$ on $v_n$ is given by the equation
\begin{equation*}
b \, v_n = c_n v_{n - 1}
\end{equation*}
for some complex constants $c_n$. It follows from the defining relations (\ref{bb}) that $c_n =
[\lambda + n]_q$. Now one can verify that the relations
\begin{gather*}
q^{\nu N} v_n = q^{\nu(\lambda + n)} v_n, \\
b^\dagger v_n = v_{n + 1}, \qquad b \, v_n = [\lambda + n]_q v_{n - 1}
\end{gather*}
endow the vector space under consideration with the structure of an $\Osc_q$-module. We denote this
module by $W^\lambda$ and the corresponding representation by $\chi^\lambda$. It is quite evident
that the modules $W^\lambda$ with $\lambda = \lambda_0 + k$ are isomorphic for all $k \in \bbZ$ and
fixed $\lambda_0$.

Now consider the $\Osc_q$-module $W^\lambda$ and assume that $[\lambda + n]_q = 0$ for some $n \in
\bbZ$. Up to an isomorphism of $\Osc_q$-modules one can assume that $n = 0$ so that $[\lambda]_q =
0$. It is the case if $\lambda = 0$ or $\lambda = \pi \rmi / \hbar$. Here we have
\begin{equation*}
b \, v_0 = 0.
\end{equation*}
It is clear that the vectors $v_n$ with $n \ge 0$ form a basis of an $\Osc_q$-submodule of
$W^\lambda$. In the case where $\lambda = 0$ we denote it by $W^+$ and the corresponding
representation by $\chi^+$. Explicit expressions for the action of the generators on the basis
vectors $v_n$, $n \in \bbZ_{\ge 0}$, of the $\Osc_q$-module $W^+$ are
\begin{gather}
q^{\nu N} v_n = q^{\nu n} v_n, \label{qoia} \\
b^\dagger v_n = v_{n + 1}, \qquad b \, v_n = [n]_q v_{n - 1}, \label{qoib}
\end{gather}
where we assume that $v_{-1} = 0$.

Let $W^-$ be a free vector space generated by vectors $v_n$, $n \in \bbZ_{\ge 0}$. One can see that
the relations
\begin{gather}
q^{\nu N} v_n = q^{- \nu (n + 1)} v_n, \label{qoiia} \\
b \, v_n = v_{n + 1}, \qquad b^\dagger v_n = - [n]_q v_{n - 1}, \label{qoiib}
\end{gather}
where we again assume that $v_{-1} = 0$, endow $W^-$ with the structure of an $\Osc_q$-module. We
denote the corresponding representation of $\Osc_q$ by $\chi^-$. One can show that this
$\Osc_q$-module is isomorphic to the quotient $\Osc_q$-module $W^0 / W^+$.

\subsubsection{Traces}

The trace on the algebra $\Osc_q$ defined with the help of the representation $\chi^\lambda$ for a
general $\lambda$ is singular. Using the representation $\chi^+$ and denoting
\begin{equation*}
\tr_+ = \tr_{\chi^+},
\end{equation*}
we see that
\begin{equation*}
\tr_+ ((b^\dagger)^{i + 1} q^{\nu N}) = 0, \qquad \tr_+ (b^{i + 1} q^{\nu N}) = 0,
\end{equation*}
and that
\begin{equation*}
\tr_+ (q^{\nu N}) = \frac{1}{1 - q^\nu}, \\
\end{equation*}
for $|q| < 1$. For $|q| > 1$ we define the trace $\tr_+$ by analytic continuation. One can also define
\begin{equation*}
\tr_- = \tr_{\chi^-}.
\end{equation*}
however, one can easily show that $\tr_- = - \tr_+$.

\subsection{\texorpdfstring{Quantum group $\uqlsltwo$}{Quantum group Uq(L(sl2))}}

\subsubsection{Definition}

It is convenient to start with the definition of $\uqtlsltwo$. Remind that $\lsltwo$ is the loop
Lie algebra of the simple Lie algebra $\sltwo$, and $\tlsltwo$ is its standard central extension,
see, for example, the book by Kac \cite{Kac90}.

The Cartan subalgebra of $\tlsltwo$ is
\begin{equation*}
\widetilde \gothh = \bbC H \oplus \bbC c,
\end{equation*}
where $H$ is the standard Cartan element of $\gothsl_2$ and $c$ the central element. Define the
Cartan elements
\begin{equation*}
h_0 = c - H, \qquad h_1 = H
\end{equation*}
so that one has
\begin{equation*}
\widetilde \gothh = \bbC h_0 \oplus \bbC h_1.
\end{equation*}
The simple positive roots $\alpha_0, \alpha_1 \in \widetilde \gothh^*$ are given by the equations
\begin{equation*}
\alpha_j(h_i) = a_{ij},
\end{equation*}
where
\begin{equation*}
(a_{ij}) = \left(\begin{array}{rr}
2 & - 2 \\
-2 & 2
\end{array} \right).
\end{equation*}

Let, as before, $\hbar$ be a complex number, such that $q = \exp \hbar \ne 0, \pm 1$. The quantum
group $\uqtlsltwo$  is a $\bbC$-algebra generated by the elements $e_i$, $f_i$, $i = 0, 1$, and
$q^x$, $x \in \widetilde \gothh$, with the relations
\begin{gather}
q^0 = 1, \qquad q^{x_1} q^{x_2} = q^{x_1 + x_2}, \label{qx} \\
q^x e_i q^{-x} = q^{\alpha_i(x)} e_i, \qquad q^x f_i q^{-x} = q^{-\alpha_i(x)} f_i, \label{qxe} \\
[e_i, f_j] = \kappa_q^{-1} \delta_{ij} \, (q^{h_i} - q^{-h_i})
\end{gather}
satisfied for all $i$ and $j$, and the Serre relations
\begin{gather}
e_i^3 e_j^{\mathstrut} - [3]_q  e_i^2 e_j^{\mathstrut} e_i^{\mathstrut}
+ [3]_q e_i^{\mathstrut} e_j^{\mathstrut} e_i^2 - e_j^{\mathstrut} e_i^3 = 0, \label{sre} \\
f_i^3 f_j^{\mathstrut} - [3]_q  f_i^2 f_j^{\mathstrut} f_i^{\mathstrut}
+ [3]_q f_i^{\mathstrut} f_j^{\mathstrut} f_i^2 - f_j^{\mathstrut} f_i^3 = 0 \label{srf}
\end{gather}
satisfied for all distinct $i$ and $j$.

The quantum group $\uqlsltwo$ can be defined as the quotient algebra of the quantum group
$\uqtlsltwo$ by the two-sided ideal generated by the elements of the form $q^{\nu c} - 1$, $\nu \in
\bbC^\times$. In terms of generators and relations the quantum group $\uqlsltwo$ is a
$\bbC$-algebra generated by the elements $e_i$, $f_i$, $i = 0, 1$, and $q^x$, $x \in \widetilde
\gothh$, with the relations (\ref{qx})--(\ref{srf}) and the additional relation\footnote{Note that
$h_0 + h_1 = c$, so that $q^{\nu(h_0 + h_1)} = q^{\nu c}$.}
\begin{equation}
q^{\nu(h_0 + h_1)} = 1. \label{qh0h1}
\end{equation}

The quantum group $\uqlsltwo$ is a Hopf algebra with the comultiplication $\Delta$ defined by the relations
\begin{gather}
\Delta(q^x) = q^x \otimes q^x, \label{dqx} \\
\Delta(e_i) = e_i \otimes 1 + q^{-h_i} \otimes e_i, \qquad \Delta(f_i)
= f_i \otimes q^{h_i} + 1 \otimes f_i, \label{defi}
\end{gather}
and with the correspondingly defined counit and antipode.

Below we denote the standard Borel subalgebras of the Lie algebra $\lsltwo$ by $\gothb_+$ and
$\gothb_-$. The Borel subalgebra $U_q(\gothb_+)$ is the subalgebra generated by $e_0$, $e_1$ and
$q^x$, $x \in \widetilde \gothh$. The Borel subalgebra $U_q(\gothb_-)$ is generated by $f_0$, $f_1$
and $q^x$, $x \in \widetilde \gothh$.

\subsubsection{Jimbo's homomorphism and universal transfer operators}

Following Jimbo~\cite{Jim86a}, we define a homomorphism
\begin{equation*}
\varphi: \uqlsltwo \to \uqsltwo
\end{equation*}
by the equations
\begin{gather*}
\varphi(q^{\nu h_0}) = q^{- \nu H}, \qquad \varphi(e_0) = F, \qquad \varphi(f_0) = E, \\
\varphi(q^{\nu h_1}) = q^{\nu H}, \qquad \varphi(e_1) = E, \qquad \varphi(f_1) = F.
\end{gather*}
Let $\widetilde \pi^{\mu}$ be the highest weight infinite-dimensional representation of $\uqsltwo$
with highest weight $\mu$ described above. We define a representation $\widetilde \varphi^{\mu}$ of
$\uqlsltwo$ as
\begin{equation*}
\widetilde \varphi^{\mu} = \widetilde \pi^{\mu} \circ \varphi.
\end{equation*}
Slightly abusing notation, we denote the corresponding $\uqlsltwo$-module by~$\widetilde V^\mu$ and
the representation by $\widetilde \pi^\mu$. We see that for this module one has
\begin{align}
&q^{\nu h_0} \, v_n = q^{- \nu (\mu - 2 n)} \, v_n, &&
q^{\nu h_1} \, v_n = q^{\nu (\mu - 2 n)} \, v_n, \label{Vh} \\
&e_0 \, v_n = v_{n + 1}, && e_1 \, v_n = [n]_q [\mu - n + 1]_q \, v_{n - 1}, \label{Ve} \\
&f_0 \, v_n = [n]_q [\mu - n + 1]_q \, v_{n - 1}, && f_1 \, v_n = v_{n + 1}. \label{Vf}
\end{align}
In the case when $\mu$ equals a non-negative integer $m$ we again abuse notation and denote the
corresponding $\uqlsltwo$-module and representation by $V^m$ and $\pi^m$.

To introduce the spectral parameter we endow $\uqlsltwo$ with a $\bbZ$-gradation assuming that the
generators $q^x$, $x \in \widetilde \gothh$, belong to the zero-grade subspace, the generators
$e_i$ belong to the subspaces with the grading indices $s_i$, and the generators $f_i$ belong to
the subspaces with the grading indices $- s_i$. Then for the mapping $\Phi_\nu$, defined by
equation (\ref{phinu}), we have
\begin{equation*}
\Phi_\nu(q^x) = q^x, \qquad \Phi_\nu(e_i) = \nu^{s_i} e_i, \qquad \Phi_\nu(f_i) = \nu^{-s_i} f_i.
\end{equation*}
Below we use the notation $s = s_0 + s_1$. Note that with this definition of a $\bbZ$-gradation
equation~(\ref{phiphir}) is true. It is useful to assume that the actual spectral parameter is a
complex number $u$, such that
\begin{equation}
\zeta = q^u = \rme^{\hbar u}. \label{zetazetaw}
\end{equation}
This assumption allows us to uniquely define arbitrary complex powers of $\zeta$.

Using the mapping $\Phi_\nu$, we come to the $\uqlsltwo$-module $\widetilde V_\zeta^\mu$ for which we have
\begin{align}
&q^{\nu h_0} \, v_n = q^{- \nu (\mu - 2 n)} \, v_n, && q^{\nu h_1} \, v_n = q^{\nu (\mu - 2 n)} \,
v_n, \label{vhs} \\* &e_0 \, v_n = \zeta^{s_0} v_{n + 1}, && e_1 \, v_n = \zeta^{s_1} [n]_q [\mu -
n + 1]_q \, v_{n - 1}, \label{ves} \\* &f_0 \, v_n = \zeta^{-s_0} [n]_q [\mu - n + 1]_q \, v_{n -
1}, && f_1 \, v_n = \zeta^{- s_1} v_{n + 1}. \label{vfs}
\end{align}
The corresponding representation is denoted by $\widetilde \pi^\mu_\zeta$. When $\mu$ equals a
non-negative integer $m$ we use for the corresponding finite-dimensional module and representation
the notations $V^m_\zeta$ and $\pi^m_\zeta$.

Now we define the universal monodromy operators
\begin{equation*}
\widetilde \calM_\mu(\zeta) = (\widetilde \varphi^\mu_\zeta \otimes \id) (\calR),
\qquad \calM_m(\zeta) = (\varphi^m_\zeta \otimes \id)(\calR)
\end{equation*}
and the universal transfer operators
\begin{gather*}
\widetilde \calT_\mu(\zeta) = ((\tr \otimes \id)
\circ (\widetilde \varphi^\mu_\zeta \otimes \id)) (\calR (t \otimes 1))
= ((\widetilde \tr_\mu \otimes \id) \circ (\varphi_\zeta \otimes \id)) (\calR (t \otimes 1)), \\[.5em]
\calT_m(\zeta) = ((\tr \otimes \id) \circ (\varphi^m_\zeta \otimes \id)) (\calR (t \otimes 1))
= ((\tr_m \otimes \id) \circ (\varphi_\zeta \otimes \id)) (\calR (t \otimes 1)).
\end{gather*}
Here $\mu$ is an arbitrary complex number and $m$ is a non-negative integer. From the explicit
expression for the universal $R$-matrix \cite{TolKho92} it follows that
\begin{equation*}
\calT_0 = 1.
\end{equation*}

Taking into account equation (\ref{trm}), we see that
\begin{equation*}
\calT_m(\zeta) = \widetilde \calT_m(\zeta) - \widetilde \calT_{- m - 2}(\zeta).
\end{equation*}
This equation suggests a definition for any complex number $\mu$ of the universal transfer operator
\begin{equation}
\calT_\mu(\zeta) = \widetilde \calT_\mu(\zeta) - \widetilde \calT_{- \mu - 2}(\zeta) = ((\tr_\mu
\otimes \id) \circ (\varphi_\zeta \otimes \id)) (\calR (t \otimes 1)), \label{tmu}
\end{equation}
where $\tr_\mu$ is the trace on $\uqsltwo$ defined by equation (\ref{trmu}). The universal transfer
operators $\calT_\mu(\zeta)$ possess the evident property
\begin{equation*}
\calT_{- \mu - 2}(\zeta) = - \calT_\mu(\zeta).
\end{equation*}
In particular, one has $\calT_{-1}(\zeta) = 0$.

\subsubsection{\texorpdfstring{Representations of the Borel subalgebras and universal
$Q$-operators}{Representations of the Borel subalgebras and universal Q-operators}}

As we noted above, to construct universal $L$-operators and universal $Q$-operators we need
representations of the Borel subalgebra $U_q(\gothb_+)$ which cannot be extended to representations
of the total quantum group $\uqlsltwo$. We consider two methods to obtain such representations.

First note that if $\varphi$ is a representation of $U_q(\gothb_+)$ and $\xi \in \widetilde
\gothh^*$, then the mapping $\varphi[\xi]$ defined by the equations
\begin{equation*}
\varphi[\xi](e_i) = \varphi(e_i), \qquad \varphi[\xi](q^x) = q^{\xi(x)} \varphi(q^x)
\end{equation*}
is a representation of $U_q(\gothb_+)$ called a {\em shifted representation\/}. It follows from
(\ref{qh0h1}) that we have to assume that
\begin{equation*}
\xi(h_0) = - \xi(h_1).
\end{equation*}
One can show that for $\xi \ne 0$ this representation cannot be extended to a representation of
$\uqlsltwo$. It follows from the formula for the universal $R$-matrix given by Khoroshkin and
Tolstoy \cite{TolKho92} that the universal $L$-operator defined with the help of the representation
$\varphi[\xi]$ is connected with the universal monodromy operator defined with the help of the
representation $\varphi$ by the relation
\begin{equation}
\calL_{\varphi[\xi]}(\zeta) = \calM_\varphi(\zeta) \, q^{\xi(h_1) (h_1 + 2 \phi) / 2}. \label{lshift}
\end{equation}
Here and below we assume that the twist element is of the form
\begin{equation}
t = q^{\phi h_1}, \label{t}
\end{equation}
where $\phi$ is a complex number. As follows from (\ref{dqx}) the element $t$ is group-like. We see
that the use of shifted representations does not give anything really new.

Let us now start with the representation $\widetilde \varphi^\mu_\zeta$ and try to consider the
limit $\mu \to \infty$. Looking at relations (\ref{vhs}) we see that we cannot take it directly for
$\widetilde \varphi^\mu_\zeta$. Therefore, we consider first a shifted representation $\widetilde
\varphi^\mu_\zeta[\xi]$ of $U_q(\gothb_+)$ for which we have
\begin{align}
&q^{\nu h_0} v_n = q^{- \nu(\mu - 2n - \xi(h_0))} v_n, && q^{\nu h_1} v_n
= q^{\nu(\mu - 2n + \xi(h_1))}  v_n, \label{shqh} \\
&e_0 v_n = \zeta^{s_0} v_{n + 1}, &&
e_1 v_n = \zeta^{s_1} [n]_q [\mu - n + 1]_q v_{n - 1}. \label{she}
\end{align}
Assume that
\begin{equation}
\xi(h_0) = - \xi(h_1) = \mu, \label{xip}
\end{equation}
and introduce a new basis
\begin{equation*}
w_n = q^{- n (\mu + 1) s_0 / s} v_n.
\end{equation*}
The relations (\ref{shqh}) take the form
\begin{equation*}
q^{\nu h_0} w_n = q^{2 \nu n} w_n, \qquad  q^{\nu h_1} w_n = q^{- 2 \nu n}  w_n,
\end{equation*}
and instead of (\ref{she}) we have
\begin{align*}
&e_0 w_n = (q^{(\mu + 1)/s} \zeta)^{s_0} \, w_{n + 1}, \\
&e_1 w_n = (q^{(\mu + 1)/s} \zeta)^{s_1} \kappa_q^{-1} (q^{- n} - q^{- 2 \mu + n - 2}) [n]_q w_{n - 1}.
\end{align*}
Denote by $\rho_\zeta^{+, \, \mu}$ the representation of $U_q(\gothb_+)$ determined by the equations
\begin{align*}
&q^{\nu h_0} v_n = q^{2 \nu n} v_n, && q^{\nu h_1} v_n = q^{- 2 \nu n}  v_n, \\
&e_0 v_n = \zeta^{s_0} \, v_{n + 1}, && e_1 v_n = \zeta^{s_1} \kappa_q^{-1} (q^{- n} - q^{- 2 \mu +
n - 2}) [n]_q v_{n - 1},
\end{align*}
and by $\rho^+_\zeta$ its limit as $\mu \to \infty$ given by the relations
\begin{align}
&q^{\nu h_0} v_n = q^{2 \nu n} v_n, && q^{\nu h_1} v_n = q^{- 2 \nu n}  v_n, \label{aria} \\
&e_0 v_n = \zeta^{s_0} \, v_{n + 1}, && e_1 v_n = \zeta^{s_1} \kappa_q^{-1} q^{- n} [n]_q v_{n - 1}. \label{arib}
\end{align}
The used notation is justified below where we consider an interpretation in terms of $q$-oscillators.

It is clear that there is an isomorphism
\begin{equation*}
\rho^{+, \, \mu}_\zeta \cong \widetilde \varphi_{q^{-(\mu + 1)/s} \zeta}^\mu[\xi],
\end{equation*}
where $\xi$ is defined by (\ref{xip}), and if we define a universal $Q$-operator by the equation
\begin{equation}
\calQ(\zeta) = \zeta^{s h_1/4} ((\tr \otimes \id)
\circ (\rho^+_\zeta \otimes \id))(\calR(t \otimes 1)), \label{defq}
\end{equation}
we have the equation
\begin{equation*}
\calQ(\zeta) = \zeta^{s h_1/4} \lim_{\mu \to \infty}
\left(\widetilde \calT_\mu(q^{-(\mu + 1)/s} \zeta) q^{-\mu(h_1 + 2 \phi)/2}\right).
\end{equation*}
A few remarks on the definition of $\calQ(\zeta)$ are in order.

The element $\zeta^{s h_1 / 4}$ is introduced to have a simple form of the universal
$TQ$-relations. In fact, this element is defined as
\begin{equation*}
\zeta^{s h_1 / 4} = q^{u s h_1 / 4},
\end{equation*}
see (\ref{zetazetaw}). Since the elements $q^{\nu h_1}$, $\nu \in \bbC$, are invertible group-like
elements commuting with the twist element (\ref{t}), they commute with the universal transfer
matrices and $Q$-operators, see sections \ref{s:ut} and \ref{s:uq}.

One can also consider the limit $\mu \to -\infty$. Here one defines the mapping $\xi$ by the relations
\begin{equation}
\xi(h_0) = \mu + 2, \qquad \xi(h_1) = - \mu - 2 \label{xim}
\end{equation}
and introduces a new basis in $\widetilde V^\mu_\zeta$ given by the equation
\begin{equation*}
w_n = \kappa_q^n q^{-n(n + 1)/2} q^{n (\mu + 1) s_0 / s} v_n.
\end{equation*}
The relations (\ref{shqh}) take now the form
\begin{equation*}
q^{\nu h_0} w_n = q^{2 \nu (n + 1)} w_n, \qquad q^{\nu h_1} w_n = q^{- 2 \nu(n + 1)}  w_n,
\end{equation*}
and instead of (\ref{she}) we have
\begin{align*}
&e_0 w_n = (q^{-(\mu + 1)/s} \zeta)^{s_0} \kappa_q^{-1} q^{n + 1} w_{n + 1}, \\
&e_1 w_n = - (q^{-(\mu + 1)/s} \zeta)^{s_1} (1 - q^{2(\mu - n + 1)}) [n]_q w_{n - 1}.
\end{align*}
Denote by $\overline \rho^{-, \, \mu}_\zeta$ the representation of $U_q(\gothb_+)$ determined by
the equations
\begin{align*}
&q^{\nu h_0} v_n = q^{2 \nu (n + 1)} v_n, && q^{\nu h_1} v_n = q^{- 2 \nu(n + 1)}  v_n, \\
&e_0 v_n = \zeta^{s_0} \kappa_q^{-1} q^{n + 1} v_{n + 1}, && e_1 v_n = - \zeta^{s_1}
(1 - q^{2(\mu - n + 1)}) [n]_q v_{n - 1},
\end{align*}
and by $\overline \rho^-_\zeta$ its limit as $\mu \to -\infty$ given by the relations
\begin{align}
&q^{\nu h_0} v_n = q^{2 \nu (n + 1)} v_n, && q^{\nu h_1} v_n = q^{- 2 \nu(n + 1)}  v_n, \label{ariia} \\
&e_0 v_n = \zeta^{s_0} \kappa_q^{-1} q^{n + 1} v_{n + 1}, && e_1 v_n = -\zeta^{s_1} [n]_q v_{n - 1}. \label{ariib}
\end{align}

There is an evident isomorphism
\begin{equation*}
\overline \rho^{-, \, \mu}_\zeta \cong \widetilde \varphi_{q^{(\mu + 1)/s} \zeta}^\mu[\xi]
\end{equation*}
with $\xi$ defined by (\ref{xim}). Introducing a new universal $Q$-operator
\begin{equation}
\overline \calQ(\zeta) = \zeta^{- s h_1/4} ((\tr \otimes \id)
\circ (\overline \rho^-_\zeta \otimes \id))(\calR(t \otimes 1)), \label{defqb}
\end{equation}
we see that
\begin{equation*}
\overline \calQ(\zeta) = \zeta^{-s h_1/4} \lim_{\mu \to - \infty}
\left(\widetilde \calT_\mu(q^{(\mu + 1)/s} \zeta) q^{-(\mu + 2)(h_1 + 2 \phi)/2}\right).
\end{equation*}

It is instructive to compare the consideration given in the present section with the formulae of
the paper \cite{BazLukMenSta10}.

\subsubsection{\texorpdfstring{Interpretation in terms of
$q$-oscillators}{Interpretation in terms of q-oscillators}}

Return again to relations (\ref{aria}) and (\ref{arib}) describing the representation
$\rho^+_\zeta$. Assume that the operators $q^{\nu N}$, $b^\dagger$ and $b$ act in the
representation space in accordance with (\ref{qoia}) and (\ref{qoib}). This allows us to write
(\ref{aria}) and (\ref{arib}) as
\begin{align*}
&q^{\nu h_0} v_n = q^{2 \nu N} v_n, && q^{\nu h_1} v_n = q^{- 2 \nu N}  v_n, \\
&e_0 v_n = \zeta^{s_0} b^\dagger v_n, && e_1 v_n = \zeta^{s_1} \kappa_q^{-1} b \, q^{-N} v_n.
\end{align*}
These equations suggest a homomorphism $\rho: U_q(\gothb_+) \to \Osc_q$ defined by
\begin{align*}
&\rho(q^{\nu h_0}) = q^{2 \nu N}, && \rho(q^{\nu h_1}) = q^{- 2 \nu N}, \\
&\rho(e_0) = b^\dagger, && \rho(e_1) = \kappa_q^{-1} b \, q^{-N}.
\end{align*}
Using the representations $\chi^+$ and $\chi^-$ of $\Osc_q$, we can now define the representations
\begin{equation*}
\rho^+ = \chi^+ \circ \rho, \qquad \rho^- = \chi^- \circ \rho
\end{equation*}
of the Borel subalgebra $U_q(\gothb_+)$. We denote the $U_q(\gothb_+)$-modules corresponding to the
representations $\rho^+_\zeta$ and $\rho^-_\zeta$ by $W^+_\zeta$ and $W^-_\zeta$. It is easy to see
that relations (\ref{aria}) and (\ref{arib}) describe the representation $\rho^+_\zeta$ as it
should be in accordance with the notation used.

It is evident that the equations
\begin{gather*}
\sigma(h_0) = h_1, \qquad \sigma(h_1) = h_0, \\
\sigma(e_0) = e_1, \qquad \sigma(e_1) = e_0, \qquad \sigma(f_0) = f_1, \qquad \sigma(f_1) = f_0
\end{gather*}
define an automorphism of $\uqlsltwo$ and, via the restriction, an automorphism of $U_q(\gothb_+)$.
Therefore, the mapping
\begin{equation*}
\overline \rho = \rho \circ \sigma
\end{equation*}
is a homomorphism from $\uqlbp$ to $\Osc_q$, and the mappings
\begin{equation*}
\overline \rho^+ = \chi^+ \circ \overline \rho, \qquad \overline \rho^- = \chi^- \circ \overline \rho
\end{equation*}
are representations of $U_q(\gothb_+)$. We denote the $U_q(\gothb_+)$-modules corresponding to the
representations $\overline \rho^+_\zeta$ and $\overline \rho^-_\zeta$ by $\overline W^+_\zeta$ and
$\overline W^-_\zeta$. One can be convinced that relations (\ref{ariia}) and (\ref{ariib}) describe
the representation $\overline \rho^-_\zeta$.

\subsubsection{\texorpdfstring{On generalized $Q$-operators}{On generalized Q-operators}}

The authors of the paper \cite{RosWes02} introduced the so called generalized $Q$-operators. To
this end they tried to find more general representations of $U_q(\gothb_+)$. The idea was to
consider a free vector space generated by vectors $u_n$, $n \in \bbZ$, and to use the ansatz
\begin{align*}
&q^{\nu h_0} u_n = q^{2 \nu n + \nu \delta} u_n, && q^{\nu h_1} u_n = q^{- 2 \nu n - \nu \delta}  u_n, \\
&e_0 u_n = \zeta^{s_0} u_{n + 1}, && e_1 u_n = \zeta^{s_1} c_n u_{n - 1},
\end{align*}
where $\delta$ and $c_n$ are some complex constants. To obtain a representation of $U_q(\gothb_+)$
one should satisfy (\ref{qx}), the first equation of (\ref{qxe}) and the Serre relations
(\ref{sre}). It is clear that only the Serre relations are not satisfied yet. To satisfy them one
has to assume that
\begin{equation*}
c_{n - 3} - [3]_q c_{n - 2} + [3]_q c_{n -1} - c_n = 0.
\end{equation*}
The general solution for this recurrence relation is
\begin{equation*}
c_n = \gamma_0 - \gamma_1 q^{- 2 n} - \gamma_2 q^{2 n},
\end{equation*}
where $\gamma_0$, $\gamma_1$ and $\gamma_2$ are arbitrary complex constants.

In a general case the trace defined with the obtained representation is singular. Let, however,
$c_n = 0$ for some $n$. Up to equivalence of representations we can assume that $c_0 = 0$, or
equivalently
\begin{equation}
\gamma_0 = \gamma_1 + \gamma_2. \label{ggg}
\end{equation}
In this case the vectors $u_n$, $n \in \bbZ_{\ge 0}$ form an invariant subspace. We denote the
corresponding $U_q(\gothb_+)$-module by $U^{\delta, \gamma_1, \gamma_2}_\zeta$.

Consider the case when $\gamma_1 \ne 0$ and $\gamma_2 \ne 0$. Here it is convenient to introduce
new parameters $\delta_1$ and $\delta_2$ such that
\begin{equation*}
\gamma_1 = q^{2 \delta_1}, \qquad \gamma_2 = q^{2 \delta_2},
\end{equation*}
and a new basis formed by the vectors
\begin{equation*}
v_n = \kappa_q^{-2 n s_0 / s} q^{- n (\delta_1 + \delta_2) s_0 / s} u_n.
\end{equation*}
One easily obtains that
\begin{equation*}
e_0 v_n = (\kappa_q^{2 / s} q^{(\delta_1 + \delta_2)/s} \zeta)^{s_0} v_{n + 1},
\quad e_1 v_n = (\kappa_q^{2 / s} q^{(\delta_1 + \delta_2)/s} \zeta)^{s_1} [n]_q
[\delta_1 - \delta_2 - n ]_q v_{n - 1}.
\end{equation*}
Remembering equation (\ref{shqh}) and (\ref{she}), we see that in the case under consideration
there is the isomorphism
\begin{equation*}
U^{\delta, \gamma_1, \gamma_2}_\zeta \cong
\widetilde V^{\delta_1 - \delta_2 - 1}_{\kappa_q^{2 / s} q^{(\delta_1 + \delta_2)/s} \zeta}[\xi]
\end{equation*}
with the element $\xi \in \widetilde \gothh^*$ determined by the equations
\begin{equation*}
\xi(h_0) = \delta + \delta_1 - \delta_2 - 1, \qquad \xi(h_1) = {} - \delta - \delta_1 + \delta_2 + 1.
\end{equation*}

Now assume that $\gamma_2 = 0$ and $\gamma_1 \ne 0$. Here $\gamma_0 = \gamma_1$, and introducing
the basis formed by the vectors
\begin{equation*}
v_n = \kappa_q^{- n s_0 / s} q^{-2 n \delta_1 s_0 / s} u_n,
\end{equation*}
we determine that
\begin{equation*}
e_0 v_n = (\kappa_q^{1 / s} q^{2 \delta_1 / s} \zeta)^{s_0} v_{n + 1}, \qquad e_1 v_n =
(\kappa_q^{1 / s} q^{2 \delta_1 / s} \zeta)^{s_1} q^{-n} [n]_q v_{n - 1}.
\end{equation*}
Having in mind equations (\ref{aria}) and (\ref{arib}), we conclude that
\begin{equation*}
U^{\delta, \gamma_1, 0}_\zeta \cong W^+_{\kappa_q^{1 / s} q^{2 \delta_1 / s} \zeta}[\xi],
\end{equation*}
where $\xi$ is defined by the equations
\begin{equation*}
\xi(h_0) = \delta, \qquad \xi(h_1) = - \delta
\end{equation*}

The last nontrivial case is when $\gamma_1 = 0$ and $\gamma_2 \ne 0$. Here defining a new basis by
the relation
\begin{equation*}
v_n = \kappa_q^n \kappa_q^{- 2 n s_0 / s} q^{-2 n \delta_2 s_0 / s} q^{-n(n + 1)/2} u_n,
\end{equation*}
we obtain
\begin{equation*}
e_0 v_n = (\kappa_q^{2 / s} q^{2 \delta_2 / s} \zeta)^{s_0} \kappa_q^{-1} q^{n + 1} v_{n + 1},
\qquad e_1 v_n = - (\kappa_q^{2 / s} q^{2 \delta_2 / s} \zeta)^{s_1} [n]_q v_{n - 1}.
\end{equation*}
Taking into account equations (\ref{ariia}) and (\ref{ariib}), we see that there is the isomorphism
\begin{equation*}
U^{\delta, 0, \gamma_2}_\zeta \cong \overline W^-_{\kappa_q^{2 / s} q^{2 \delta_2 / s} \zeta}[\xi],
\end{equation*}
where
\begin{equation*}
\xi(h_0) = \delta - 2, \qquad \xi(h_1) = {} - \delta + 2.
\end{equation*}

In fact, one can show that even in the case when (\ref{ggg}) is not satisfied there are
isomorphisms of $U_q(\gothb_+)$-modules defined in this section with $U_q(\gothb_+)$-modules
defined in sections \ref{sltwom} and \ref{oscm} for $\lambda \ne 0$. However, in this case we meet
the problem of the singularity of the trace. Thus, the generalized $Q$-operators introduced in the
paper~\cite{RosWes02} are equivalent either to usual transfer operators or to usual $Q$-operators.
Nevertheless, the additional representations considered in the present paper and in~\cite{RosWes02}
can be used to establish the integrability of some interesting quantum systems, see, for example,
\cite{Ant97}.

\section{Example. Universal functional relations}

\subsection{Commutativity relations}

It is worth to remind that since $q^{\nu h_1}$ for any $\nu \in \bbC$ is an invertible group-like
element of $\uqlsltwo$ and commutes with the twist element $q^{\phi h_1}$, it commutes with the
universal transfer operators $\widetilde \calT_\mu(\zeta)$, $\calT_\mu(\zeta)$ and with the
universal $Q$-operators $\calQ(\zeta)$, $\overline \calQ(\zeta)$.

There are functional relations which are due only to the fact that the universal transfer operators
and the universal $Q$-operators are constructed from the universal $R$-matrices. These are the
commutativity relations for the universal transfer matrices
\begin{equation*}
\boxed{[\widetilde \calT_{\mu_1}(\zeta_1), \widetilde \calT_{\mu_2}(\zeta_2)] = 0, \qquad
[\calT_{\mu_1}(\zeta_1), \calT_{\mu_2}(\zeta_2)] = 0, \qquad [\widetilde \calT_{\mu_1}(\zeta_1),
\calT_{\mu_2}(\zeta_2)] = 0} \, ,
\end{equation*}
see relation (\ref{tzetatzeta}), and the commutativity of the universal transfer operators and the
universal $Q$-operators
\begin{gather*}
\boxed{[\widetilde \calT_\mu(\zeta_1), \calQ(\zeta_2)] = 0,
\qquad [\widetilde \calT_\mu(\zeta_1), \overline \calQ(\zeta_2)] = 0} \, , \\[.5em]
\boxed{[\calT_\mu(\zeta_1), \calQ(\zeta_2)] = 0, \qquad [\calT_\mu(\zeta_1), \overline \calQ(\zeta_2)] = 0} \, ,
\end{gather*}
see relation (\ref{qzetatzeta}).

Another set of commutativity relations follows from the properties of the representations used to
define the universal transfer operators and the universal $Q$-operators. Having in mind that the
universal $Q$-operators are obtained from the universal transfers operators by  limiting procedure
and that the universal transfer operators commute, we obtain
\begin{equation*}
\boxed{[\calQ(\zeta_1), \calQ(\zeta_2)] = 0, \qquad [\calQ(\zeta_1), \overline \calQ(\zeta_2)] = 0,
\qquad [\overline \calQ(\zeta_1), \overline \calQ(\zeta_2)] = 0} \, .
\end{equation*}

\subsection{\texorpdfstring{Universal $TQ$-relations}{Universal TQ-relations}}

We see that the universal $Q$-operators $\calQ(\zeta)$ and $\overline \calQ(\zeta)$ commute for
coinciding and different values of the spectral parameters. More detailed information on their
product can be obtained from relation (\ref{qzetaqzeta}). Analysing the structure of the
representation $\rho^+_{\zeta_1} \otimes_\Delta \overline \rho^-_{\zeta_2}$ \cite{BazLukZam97,
BooJimMiwSmiTak09, BooGoeKluNirRaz12}, one can see that the $U_q(\gothb_+)$-module $W^+_{\zeta_1}
\otimes \overline W{}^-_{\zeta_2}$ has an increasing filtration
\begin{equation*}
\{0\} = (W^+_{\zeta_1} \otimes \overline W{}^-_{\zeta_2})^{\mathstrut}_{-1}
\subset (W^+_{\zeta_1} \otimes \overline W{}^-_{\zeta_2})^{\mathstrut}_0
\subset (W^+_{\zeta_1} \otimes \overline W{}^-_{\zeta_2})^{\mathstrut}_1 \subset \ldots,
\end{equation*}
where $(W^+_{\zeta_1} \otimes \overline W{}^-_{\zeta_2})^{\mathstrut}_k$ are
$U_q(\gothb_+)$-submodules with the quotient modules
\begin{equation}
(W^+_{\zeta_1} \otimes \overline W{}^-_{\zeta_2})^{\mathstrut}_k / (W^+_{\zeta_1} \otimes \overline
W{}^-_{\zeta_2})^{\mathstrut}_{k - 1} \cong \widetilde V^\mu_\zeta[\xi_k]. \label{pmqm}
\end{equation}
Here the elements $\xi_k \in \widetilde \gothh^*$ are determined by the relations
\begin{equation}
\xi_k(h_0) = \mu + 2 k + 2, \qquad \xi_k(h_1) = - \mu - 2 k - 2, \label{xik}
\end{equation}
The parameters $\zeta$ and $\mu$ are connected with the parameters $\zeta_1$ and $\zeta_2$ as
\begin{equation}
\zeta = (\zeta_1 \zeta_2)^{1/2}, \qquad q^{\mu + 1} = (\zeta_1 / \zeta_2)^{(s_0 + s_1)/2}.\label{zeta}
\end{equation}
The inverse transformation to the parameters $\zeta_1$ and $\zeta_2$ is
\begin{equation*}
\zeta_1 = q^{(\mu + 1)/s} \zeta, \qquad \zeta_2 = q^{-(\mu + 1)/s} \zeta.
\end{equation*}

It follows from relations (\ref{defq}), (\ref{defqb}), (\ref{pmqm}) and (\ref{lshift}) that
\begin{equation*}
\calQ(q^{(\mu + 1)/s} \zeta) \overline \calQ(q^{-(\mu + 1)/s} \zeta) = \widetilde \calT_\mu(\zeta)
\, \frac{q^{-(\mu + 1) \phi}}{q^{(h_1 + 2 \phi)/2} - q^{- (h_1 + 2 \phi)/2}}.
\end{equation*}
Hence, we have
\begin{equation}
\widetilde \calT_\mu(\zeta) = q^{(\mu + 1) \phi} \calC \, \calQ(q^{(\mu + 1)/s} \zeta) \overline
\calQ(q^{-(\mu + 1)/s} \zeta), \label{ttmu}
\end{equation}
where
\begin{equation*}
\calC = q^{(h_1 + 2 \phi)/2} - q^{- (h_1 + 2 \phi)/2}.
\end{equation*}

Rewriting (\ref{ttmu}) as
\begin{equation}
\widetilde \calT_\mu(q^{\nu/s} \zeta) = q^{(\mu + 1) \phi} \calC \, \calQ(q^{(\mu + \nu + 1)/s}
\zeta) \overline \calQ(q^{- (\mu - \nu + 1)/s} \zeta) \label{tctqq}
\end{equation}
and introducing new parameters
\begin{equation}
\alpha = \mu + 1 + \nu, \qquad \beta = - (\mu + 1) + \nu, \label{pab}
\end{equation}
we come to the equation
\begin{equation*}
\widetilde \calT_{(\alpha - \beta)/2 - 1}(q^{(\alpha + \beta)/2s} \zeta) = q^{(\alpha - \beta) \phi
/ 2} \calC \, \calQ(q^{\alpha/s}  \zeta) \overline \calQ(q^{\beta/s} \zeta).
\end{equation*}
Using (\ref{ttmu}), we easily determine that
\begin{equation}
q^{\gamma \phi / 2} \widetilde \calT_{(\alpha - \beta)/2 - 1}(q^{(\alpha + \beta)/2s} \zeta)
\calQ(q^{\gamma/s} \zeta) = q^{\alpha \phi / 2} \widetilde \calT_{(\gamma - \beta)/2 -
1}(q^{(\gamma + \beta)/2s} \zeta) \calQ(q^{\alpha/s} \zeta) \label{ttq}
\end{equation}
and that
\begin{equation}
q^{- \gamma \phi / 2} \widetilde \calT_{(\alpha - \beta)/2 - 1}(q^{(\alpha + \beta)/2s} \zeta)
\overline \calQ(q^{\gamma/s} \zeta) = q^{- \beta \phi / 2} \widetilde \calT_{(\alpha - \gamma)/2
- 1}(q^{(\alpha + \gamma)/2s} \zeta) \overline \calQ(q^{\beta/s} \zeta). \label{ttbq}
\end{equation}

It follows from (\ref{tmu}) and (\ref{ttmu}) that for the universal transfer operators
$\calT_\mu(\zeta)$ we have
\begin{equation}
\calT_\mu(\zeta) = \calC \Bigl[ q^{(\mu + 1) \phi} \calQ(q^{(\mu + 1)/s}  \zeta) \overline
\calQ(q^{- (\mu + 1)/s} \zeta) \\*- q^{-(\mu + 1) \phi} \calQ(q^{-(\mu + 1)/s}  \zeta) \overline
\calQ(q^{(\mu + 1)/s} \zeta) \Bigr]. \label{ctqq}
\end{equation}
In particular, for $\mu = 0$ we come to the Wronskian type relation
\begin{equation*}
\boxed{\calC \Bigl[ q^\phi \calQ(q^{1/s}  \zeta) \overline \calQ(q^{- 1/s} \zeta) - q^{-\phi}
\calQ(q^{-1/s}  \zeta) \overline \calQ(q^{1/s} \zeta) \Bigr] = 1} \, .
\end{equation*}

From (\ref{ctqq}) it is easy to obtain the equation
\begin{equation*}
\calT_{(\alpha - \beta)/2 - 1}(q^{(\alpha + \beta)/2s} \zeta) = \calC \Bigl[ q^{(\alpha - \beta)
\phi / 2} \calQ(q^{\alpha/s}  \zeta) \, \overline \calQ(q^{\beta/s} \zeta) \\- q^{(\beta - \alpha)
\phi / 2} \calQ(q^{\beta/s} \zeta) \, \overline \calQ(q^{\alpha/s} \zeta) \Bigr],
\end{equation*}
which implies that
\begin{multline}
q^{\gamma \phi / 2} \calT_{(\alpha - \beta)/2 - 1}(q^{(\alpha + \beta)/2s} \zeta)
\calQ(q^{\gamma / s} \zeta)
+ q^{\alpha \phi / 2} \calT_{(\beta - \gamma)/2 - 1}(q^{(\beta + \gamma)/2s} \zeta)
\calQ(q^{\alpha / s} \zeta) \\*[.5em]
+ q^{\beta \phi / 2} \calT_{(\gamma - \alpha)/2 - 1}(q^{(\gamma + \alpha)/2s} \zeta)
\calQ(q^{\beta / s} \zeta) = 0, \label{utqr}
\end{multline}
and that
\begin{multline}
q^{- \gamma \phi / 2} \calT_{(\alpha - \beta)/2 - 1}(q^{(\alpha + \beta)/2s} \zeta)
\overline \calQ(q^{\gamma / s} \zeta)
+ q^{-\alpha \phi / 2} \calT_{(\beta - \gamma)/2 - 1}(q^{(\beta + \gamma)/2s} \zeta)
\overline \calQ(q^{\alpha / s} \zeta) \\*[.5em]
+ q^{-\beta \phi / 2} \calT_{(\gamma - \alpha)/2 - 1}(q^{(\gamma + \alpha)/2s} \zeta)
\overline \calQ(q^{\beta / s} \zeta) = 0. \label{utbqr}
\end{multline}
We call the equations (\ref{utqr}) and (\ref{utbqr}) the {\em universal $TQ$-relations\/}. Putting
\begin{equation*}
\alpha = \gamma - 2, \qquad \beta = \gamma + 2,
\end{equation*}
we obtain the relations of more usual form,
\begin{gather*}
\boxed{\calT(\zeta) \calQ(\zeta) = q^\phi \calQ(q^{2 / s} \zeta) + q^{-\phi} \calQ(q^{- 2 / s}
\zeta)} \, ,  \\*[.5em] \boxed{\calT(\zeta) \overline \calQ(\zeta) = q^{-\phi} \overline \calQ(q^{2
/ s} \zeta) + q^\phi \overline \calQ(q^{- 2 / s} \zeta)} \, ,
\end{gather*}
where we denote $\calT(\zeta) = \calT_1(\zeta)$.

\subsection{\texorpdfstring{Universal $TT$-relations}{Universal TT-relations}}

Using relation (\ref{ttmu}), we obtain from (\ref{ttq}), or from~(\ref{ttbq}), the equation
\begin{multline*}
\widetilde \calT_{(\alpha - \beta)/2 - 1}(q^{(\alpha + \beta)/2s} \zeta) \widetilde \calT_{(\gamma
- \delta)/2 - 1}(q^{(\gamma + \delta)/2s} \zeta) \\= \widetilde \calT_{(\gamma - \beta)/2 -
1}(q^{(\gamma + \beta)/2s} \zeta) \widetilde \calT_{(\alpha - \delta)/2 - 1}(q^{(\alpha +
\delta)/2s} \zeta).
\end{multline*}
For the universal transfer operators $\calT_\mu(\zeta)$ defined by (\ref{tmu}) we obtain
\begin{multline*}
\calT_{(\alpha - \beta)/2 - 1}(q^{(\alpha + \beta)/2s} \zeta) \calT_{(\gamma - \delta)/2 -
1}(q^{(\gamma + \delta)/2s} \zeta) \\*[.5em] = \calT_{(\alpha - \gamma)/2 - 1}(q^{(\alpha +
\gamma)/2s} \zeta) \calT_{(\beta - \delta)/2 - 1}(q^{(\beta + \delta)/2s} \zeta) \\*[.5em]-
\calT_{(\beta - \gamma)/2 - 1}(q^{(\beta + \gamma)/2s} \zeta) \calT_{(\alpha - \delta)/2 -
1}(q^{(\alpha + \delta)/2s} \zeta).
\end{multline*}
We call these relations the {\em universal $TT$-relations\/}. There are two interesting special
cases of these relations. In the first case we put
\begin{equation*}
\alpha = \gamma + 2, \qquad \beta = \delta + 2
\end{equation*}
and obtain
\begin{equation*}
\boxed{\calT_\mu(q^{1/s} \zeta) \calT_\mu(q^{-1/s} \zeta)
= 1 + \calT_{\mu - 1}(\zeta) \calT_{\mu + 1}(\zeta)} \, ,
\end{equation*}
where $\mu = (\gamma - \delta) / 2 - 1$. In the second case we put
\begin{equation*}
\alpha = \gamma + 2, \qquad \beta = \gamma - 2
\end{equation*}
and obtain
\begin{equation*}
\boxed{\calT(\zeta) \calT_\mu(q^{-(\mu + 1)/s} \zeta) = \calT_{\mu + 1}(q^{- \mu/s} \zeta) +
\calT_{\mu - 1}(q^{-(\mu + 2)/s} \zeta)} \,,
\end{equation*}
where again $\mu = (\gamma - \delta)/2 - 1$.

\section{Conclusion}

We gave and discussed general definitions and facts on the application of quantum groups to the
construction of functional relations in the theory of integrable systems. As an example, we
reconsidered the case of the quantum group $\uqlsltwo$ related to the six-vertex model and the XXZ
spin chain. We gave the full set of the functional relations in the form independent of the
representation of the quantum group in the quantum space. The specialization of the universal
$TQ$-relations and universal $TT$-relations to the case of the isotropic six-vertex model is
obtained by other methods in the papers \cite{DerMan06, BazLukMenSta10}.

\subsubsection*{Acknowledgements} This work was supported in part by the Volkswagen Foundation.
Kh.S.N. was supported in part by the grant NS-5590.2012.2. A.V.R. was supported in part by the RFBR
grant \#~10-01-00300. He would like to thank the Max Planck Institute for Mathematics in Bonn for
the hospitality extended to him during his stay in February-May 2012.




\newcommand{\noopsort}[1]{}
\providecommand{\bysame}{\leavevmode\hbox to3em{\hrulefill}\thinspace}
\providecommand{\href}[2]{{#2}}

\end{document}